\documentclass{mem}
\usepackage{natbib}\usepackage{txfonts}\usepackage{balance}
\usepackage{graphicx}
\usepackage[a4paper]{hyperref}
\usepackage{verbatim}
\idline{76/4, eds. A.R. Walker \& G. Bono}{1}
\begin{document}
\def\teff{$T\rm_{eff }$}
\def\kms{$\mathrm {km s}^{-1}$}
\def\logg{$\log{g}$}
\def\sun{\hbox{$\odot$}}
%
%
%
%
\title{Exciting new features in the frequency spectrum
  of the EC\,14026 star HS\,0702+6043}

\subtitle{Simultaneous g-modes and p-modes in a sdB pulsator}

\author{
  S.~Schuh\inst{1} 
  \and
  J.~Huber\inst{1} \and  S.~Dreizler\inst{1} \and  E.M.~Green\inst{2} \and
  T.~Stahn\inst{1} \and  S.~Randall\inst{3} \and
  T.-O.~Husser\inst{1} \and  U.~Heber\inst{4} \and  S.~O'Toole\inst{4} \and
  G.~Fontaine\inst{3}
}

\offprints{S.~Schuh}

\institute{
  Inst. f\"ur Astrophysik, Univ.\,G\"ottingen, Friedrich-Hund-Pl.\,1,
  37077\,G\"ottingen, Germany
  \and
  Steward Observatory, Univ.\ of Arizona, 933\,N\,Cherry~Ave., Tucson~AZ\,85721-0065, USA \and
  D{\'e}partement de physique, Univ.\ de Montr{\'e}al,
  CP~6128 Montr{\'e}al, Canada, H3C-3J7 \and
  Remeis-Sternwarte, Univ.\,Erlangen-N\"urnberg, Sternwartstr.\,7,
  96049\,Bamberg, Germany
}

\authorrunning{Schuh}

\titlerunning{Simultaneous g-modes and p-modes in the EC\,14026 star HS\,0702$+$6043}

\abstract{
  The discovery of a long-period g-mode oscillation in the previously
  known short-period p-mode sdB pulsator HS\,0702$+$6043 makes this star
  an extraordinary object, unique as a member of the family of sdB
  pulsators, and one of the very few known pulsating stars overall
  exhibiting excited modes along both the acoustic and gravity
  branches of the nonradial pulsation spectrum. Because p-modes and
  g-modes probe different regions of a pulsating star, HS\,0702$+$6043
  holds a tremendous potential for asteroseismological
  investigations. We present preliminary results from the first
  extended campaign on this object.
}
\maketitle{}
\section{The two classes of sdB pulsators}
Subdwarf B stars populate the extreme horizontal branch (EHB) in the
effective temperature range of 22\,000 to 40\,000\,K and have surface
gravity values from \logg=5.0 to 6.2 in cgs units. The masses of
these hot, evolved objects cluster around 0.5~M$_{\sun}$. They are
believed to be core helium-burning but with hydrogen envelopes too
thin to sustain H-shell burning. Standard tracks of stellar evolution
do not cross the EHB region since they do not produce inert hydrogen
envelopes; the evolutionary history of sdBs largely remains a puzzle.
\par 
SdB internal structure might be investigated through the multi-mode
oscillations exhibited by a fraction of sdB stars. One sdBV group
known as EC\,14026 variables shows short-period p-mode pulsations with
periods of a few minutes and amplitudes of a few tens mmag. The more
recently discovered \mbox{lpsdBV} (long-period sdB variables, the
prototype is PG\,1716$+$426) pulsate in g-modes of about 30-80~min at
mmag amplitudes.
A $\kappa$ mechanism drives the low-order p-modes, where the required
opacity bump is due to iron accumulated by diffusion. 
The same $\kappa$ mechanism can also drive high-order g-modes, so that
the sdBV\,/\,\mbox{lpsdBV} groups might
have a main sequence analogy in the $\beta$\,Cep\,/\,[SPB] variables
(high\,/\,low frequency, high\,/\,low temperature).
Theoretical aspects are discussed in more detail by Charpinet et al.;
Fontaine et al.; and Randall et al.\ (all these proceedings).
\section{The hybrid object HS\,0702+6043}
In a 1998\,/\,1999 search program for sdB variables, we discovered the
m$_B$=15 star HS\,0702+6043 as one of the coolest, high-amplitude sdBV
pulsators \citep{2002A&A...386..249D}. Its stellar parameters
(28\,400\,K, \logg=5.35) place HS\,0702$+$6043 at the cool end of the
EC\,14026 instability region.
\par 
We subsequently demonstrated the existence of an additional
photometric variation at a much longer time-scale of 1\,h. This
initial discovery was confirmed by further data obtained under
excellent observing conditions during two nights in February 2004.
The detection suggested a complex nature of the amplitude structure in
the low-frequency range, excluding alternative explanations such as
binarity or rotation at a high level of confidence. Due to this 
clear pulsational character of the variations, HS\,0702$+$6043 must be
regarded as a member of both classes and represents the first hybrid
object that exhibits both p- and g-mode pulsations
\citep{2005ewwd.conf....2S}.  To resolve the sub-structure in the
low-frequency range, a dedicated 10-day coordinated campaign was
organised in January 2005 at the 2.2\,m Calar Alto (Spain) and the
1.55\,m Steward telescope on Mount Bigelow (U.S.A.).
\par
The dominant feature in the frequency spectrum is a 2754\,$\mu$Hz
(363\,s, $f_1$) pulsation at an amplitude of $\approx$\,22-29\,mmag. A
second short period at 2606\,$\mu$Hz (383\,s, $f_2$) at a much smaller
amplitude of $\approx$\,4\,mmag was suspected in the 1999 data, and confirmed in
the 2004 and 2005 data. The value published for $f_2$ in
\citet{2002A&A...386..249D} is separated by one daily alias from the
current determination.
\par
In the longer-period range, a peak near $\approx$~1\,h is marginally
significant in the 1999 data, and clearly detected in the 2004 data.
There is residual power above the $3 \sigma$ detection level,
determined from false alarm probability analysis, after prewhitening
of one sinusoidal period in the low-frequency range from the 2004
data.  This complex structure can either be attributed to the phase
discontinuities and amplitude variability (a quasi-periodic behaviour
characterstic of many of the known long-period sdB pulsators) or, when
the amplitude variability is interpreted as beating, suggests the
presence of unresolved frequencies.  In the 2005 data, the highest
peak in the low-frequency range ($f_3$=271$\mu$Hz, or 3690\,s) is
consistent with the previous detection, and it is, at the better
frequency resolution now available, \emph{no longer consistent} with a
value of \mbox{$2(f_1-f_2)$}, as suggested previously. 
The new data are generally much noisier,
with significant very-low frequency contributions due to atmospheric
instability complicating the analysis, and require further work.
\section{Overlapping instability regions}
The identification of HS\,0702$+$6043 as a member of \emph{both}
classes of sdB pulsators simultaneously implies that the instability
regions overlap. One important consequence of this is that other such
hybrid objects, which are of considerable interest in asteroseismology
since the different modes probe different regions of a pulsating star,
might exist. And indeed, following the initial announcement of
HS\,0702$+$6043 \citep{2005ewwd.conf....2S}, a second such object has
been published (Balloon~090100001:
\citealt{2004A&A...418..243O}\,/\,\citealt{2005MNRAS.360..737B}). 
\par
It is interesting to note that a hybrid $\delta$~Sct and $\gamma$~Dor
object has also been found, following the prediction of the likely
existence of such stars with both $\delta$~Sct p-modes and
$\gamma$~Dor g-modes oscillations by Dupret et al.\ (these
proceedings).
\end{document}